\documentclass[journal]{IEEEtran}

\ifCLASSINFOpdf
\else
\fi

\usepackage[colorlinks = true, linkcolor=red,
    filecolor=blue,
    urlcolor=blue,
    citecolor=green,
    ]{hyperref}
\usepackage{footmisc}
\usepackage{enumitem}
\usepackage{graphicx}
\usepackage{multirow}
\usepackage{amsmath}
\usepackage{lineno}
\usepackage{amsfonts}
\usepackage{bm}

\usepackage{subfigure}
\usepackage{color}
\usepackage{booktabs}
\usepackage[misc]{ifsym}
\usepackage{cite}
\usepackage{url}
\begin{document}
%
\title{TransAttUnet: Multi-level Attention-guided U-Net with Transformer for Medical Image Segmentation}
%
%
%

\author{Bingzhi~Chen,~\IEEEmembership{}
        Yishu~Liu,~\IEEEmembership{}
        Yingjian Li,~\IEEEmembership{}
        Zheng~Zhang,~\IEEEmembership{Senior Member,~IEEE,}\\
        Guangming~Lu,~\IEEEmembership{Member,~IEEE,}
        and Adams Wai Kin Kong,~\IEEEmembership{Member,~IEEE.}


\thanks{\copyright \copyright 2022 IEEE. Personal use of this material is permitted. Permission from IEEE must be obtained for all other uses, in any current or future media, including reprinting/republishing this material for advertising or promotional purposes, creating new collective works, for resale or redistribution to servers or lists, or reuse of any copyrighted component of this work in other works.}
\thanks{B. Chen is with the Shenzhen Medical Biometrics Perception and Analysis Engineering Laboratory, Harbin Institute of Technology, Shenzhen $518055$, China, and also with the School of Computer Science and Engineering, Nanyang Technological University, Singapore $639798$. (e-mail: chenbingzhi.smile@gmail.com)}
\thanks{Y. Liu, Y. Li, Z. Zhang, and G. Lu are with the Shenzhen Medical Biometrics Perception and Analysis Engineering Laboratory, Harbin Institute of Technology, Shenzhen $518055$, China. (e-mail: liuyishu.smile@gmail.com, hit\_lyj@126.com, darrenzz219@gmail.com, luguangm@hit.edu.cn)}
\thanks{A. Kong is with the School of Computer Science and Engineering, Nanyang Technological University, Singapore, $639798$. (e-mail: AdamsKong@ntu.edu.sg)}
}

\markboth{IEEE Transactions on Instrumentation \& Measurement,~Vol.~x, No.~x, ~2022}%
{Shell \MakeLowercase{\textit{et al.}}: Bare Demo of IEEEtran.cls for IEEE Journals}
%



\maketitle

\begin{abstract}
Accurate segmentation of organs or lesions from medical images is crucial for reliable diagnosis of diseases and organ morphometry. In recent years, convolutional encoder-decoder solutions have achieved substantial progress in the field of automatic medical image segmentation.
Due to the inherent bias in the convolution operations, prior models mainly focus on local visual cues formed by the neighboring pixels, but fail to fully model the long-range contextual dependencies.
In this paper, we propose a novel Transformer-based
 Attention Guided Network called \textit{TransAttUnet}$\footnote{The preprint of our work has been posted at: \url{ https://doi.org/10.48550/arXiv.2107.05274}. Our code and pre-trained models are available at: \url{https://github.com/YishuLiu/TransAttUnet}}$, in which the multi-level guided attention and multi-scale skip connection are designed to jointly enhance the performance of the semantical segmentation architecture.
Inspired by Transformer, the self-aware attention (SAA) module with Transformer Self Attention (TSA) and Global Spatial Attention (GSA) is incorporated into TransAttUnet to effectively learn the non-local interactions among encoder features.
Moreover, we also use additional multi-scale skip connections between decoder blocks to aggregate the upsampled features with different semantic scales. In this way, the representation ability of multi-scale context information is strengthened to generate discriminative features.
Benefitting from these complementary components, the proposed TransAttUnet can effectively alleviate the loss of fine details caused by the stacking of convolution layers and the consecutive sampling operations, finally improving the segmentation quality of medical images.
Extensive experiments on multiple medical image segmentation datasets from different imaging modalities demonstrate that the proposed method consistently outperforms the state-of-the-art baselines.
\end{abstract}


\IEEEpeerreviewmaketitle

\section{Introduction}
\label{sec:1}
\begin{figure}[t]
\centering
\includegraphics[width=3.5in]{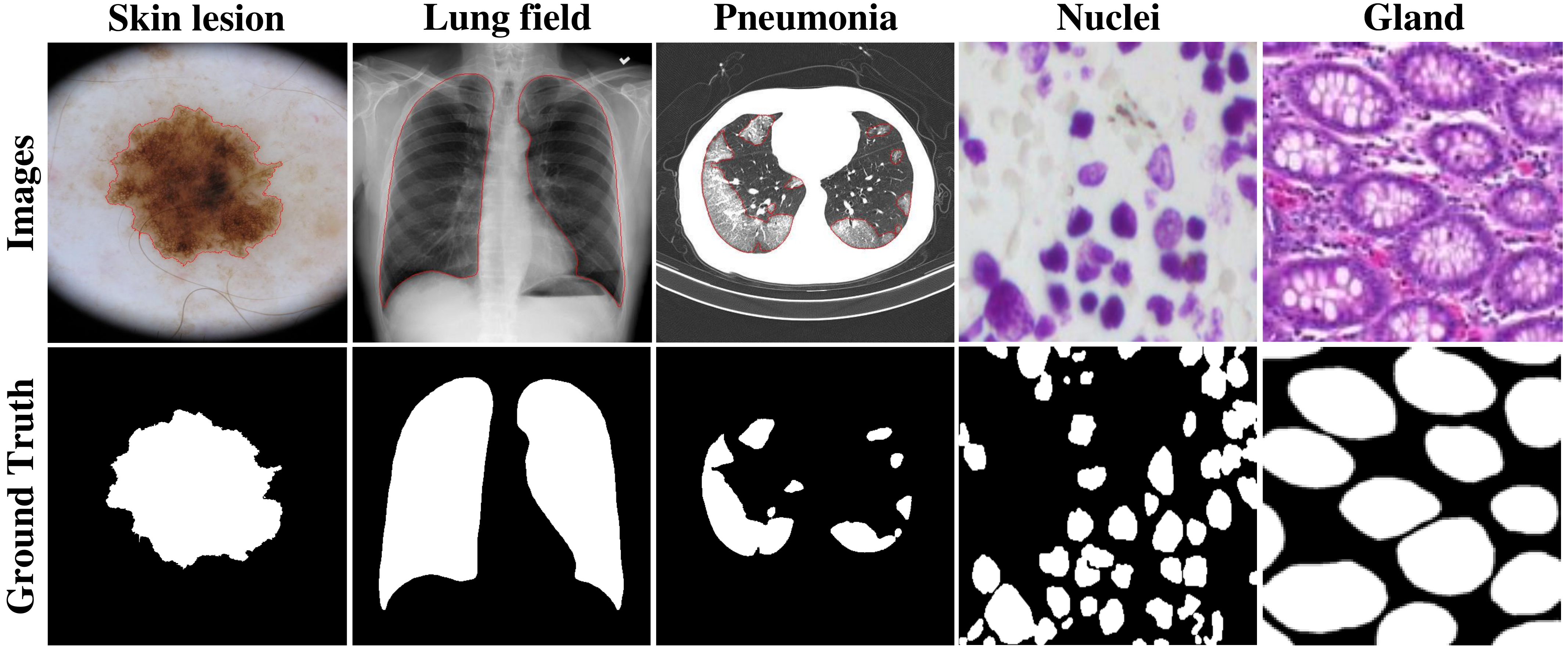}
\caption{Examples of medical images with the corresponding semantic segmentation annotations.}
\label{fig:1}
\end{figure}

\IEEEPARstart{I}{n} clinical diagnosis, the purpose of medical image segmentation is to delineate the objects of interest from the complex background on various biomedical images\cite{roy2022lwmla, pramanik2019seg, yang2020colon}, such as X-ray, Computerized Tomography (CT), Magnetic Resonance Imaging (MRI), and Ultrasound.
It is useful for quantitative diagnosis and morphological analysis of specific lesions in human organs and tissues.
As shown in Fig.~\ref{fig:1}, it requires enormous effort and patience to handle the complex contours and textures. However, traditional manual annotation heavily relies on clinical experiences. Measurements based on clinicians’ manual annotations might be very accurate but highly labor-intensive under standard clinical settings.
Therefore, it is in great demand to develop accurate medical image segmentation methods.

In recent years, deep encoder-decoder architectures have proven to be effective in recovering details of the segmented objects and gradually become the industry's facto benchmark for medical image segmentation.
The main purpose of convolutional operations used in deep encoder-decoder architectures is to extract local features of images by gathering local information from the neighboring pixels.
Typically, the stacks of convolution layers and the consecutive sampling operations constantly extend the receptive field and aggregate the global filter responses to determine the coarse object boundaries.
Benefitting from deconvolution operations \cite{fu2020contextual}, Fully Convolutional Networks (FCN) \cite{long2015fully} can naturally operate on images with various sizes and generate appropriate dimension outputs according to the corresponding inputs. Inspired by FCN, U-Net \cite{ronneberger2015u} directly combines low-level features from the analysis path and deep features in the expansion path through encoder-decoder skip connections, which can achieve the trade-off capability between local information and contextual information.

Despite the interesting design and encouraging performance, the above design limits the information flow due to the convolutional operations and becomes a bottleneck for performance enhancement.
Specifically, these architectures have several drawbacks: (1) these architectures generally lack the ability to model long-range feature dependencies due to the inherent inductive biases \cite{luo2016understanding} in the convolution computing paradigm; (2) the low-level features might be prevented from being transmitted to the subsequent convolutional layers in the process of pooling and convolution \cite{chen2019channel}, which compromise the quality of local information and degrade the semantic segmentation performance; (3) the existing skip connection mechanism is only performed on the same scale feature maps without exploring the relationship between feature maps from different decoding stages, which cannot guarantee the consistency of the feature representations and semantic embeddings \cite{zhou2018unet++}.

To solve these issues, researchers have put considerable effort in developing various variants of U-Net. The existing works follow three research lines: (1) attention-guided approaches; (2) context-based approaches; (3) Transformer-based approaches.
For example, prior attention-guided works, e.g. Attention U-Net \cite{oktay2018attention} and Channel-UNet\cite{chen2019channel}, attempted to leverage various attention mechanisms to optimize the spatial information of feature maps extracted from encoding, decoding, and output stages.
The main idea of these attention mechanisms is to generate a confidence mask to recalibrate the response of the original feature maps.
Different from the above approaches, the second one is to explore the contextual information by using multi-scale connections.
It is indisputable that both high-level abstract information and low-level pixel information are of great significance in developing accurate segmentation.
Therefore, varying degrees of context-based shortcut connection components have been embedded into the U-shaped architectures, such as Unet++ \cite{zhou2018unet++}, and  MA-Unet \cite{cai2020ma}, in order to capture broader and deeper contextual representations.
Recently, Transformer \cite{jaderberg2015spatial}\cite{parmar2018image} is getting great attention from computer vision (CV) researchers. Especially, some recent works, such as  TransUNet \cite{chen2021transunet} and MedT\cite{valanarasu2021medical}, have tried to incorporate Transformer with CNN-based model to boost the performance of medical image segmentation.
In general, Transformer-based approaches have very high computation complexities to process feature maps, and often required models pretrained on a large external dataset.
As the core of Transformer, the multi-head mechanism\cite{tao2018get} can capture the long-range contextual information by running through the scaled dot-product attention multiple times in parallel. Therefore, it is undoubtedly an ideal supplemental part that can efficiently compensate for the design flaws of U-Net.

\begin{figure*}[t]
\centering
\includegraphics[width=7.0in]{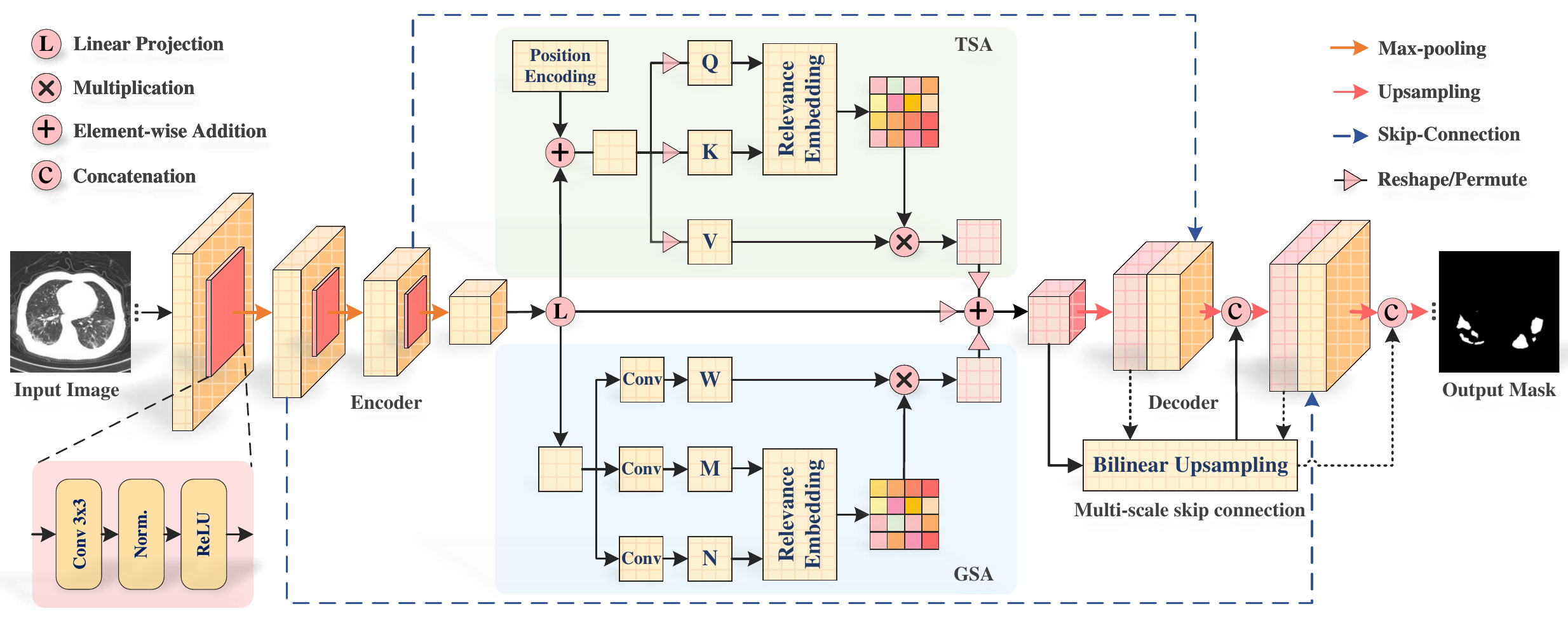}
\caption{Illustration of the proposed TransAttUnet for automatic medical image segmentation. (a) Both TSA and GSA mechanisms are embedding into the SAA module to model the long-range
interactions and global spatial relationships.
 (b) The multi-scale skip connections between decoder blocks are designed to aggregate the downsampled features of varying semantic scales by progressive upsampling, concatenation, and convolution.}
\label{fig:2}
\end{figure*}

Inspired by the aforementioned advanced works, in this paper we propose a novel multi-level attention-guided U-Net with Transformer, dubbed TransAttUnet, that can effectively enhance the segmentation accuracy of traditional U-shaped architecture by jointly utilizing multi-level guided attention and multi-scale skip connection.
The architecture of the proposed TransAttUnet is shown in Fig.~\ref{fig:2}.
Specifically, a robust self-aware attention (SAA) module is first embedded into the proposed TransAttUnet as the bridge between the encoder and decoder subnetworks.
As the key component of TransAttUnet, the purpose of the SAA module is to concurrently leverage the powerful abilities of transformer self attention (TSA) and global spatial attention (GSA) to establish effective long-range interactions and global spatial relationships between encoder semantic features.
Motivated by the idea of residual and dense shortcut connections, a multi-scale skip connection scheme is added into decoder sub-networks with a series of transition operations, including upsampling, concatenation, and convolution. Thus, it can flexibly aggregate the residual or dense contextual feature maps from decoder blocks of varying semantic scales step by step, in order to generate more discriminative feature representations.
With the contributions of these complementary components, the proposed method can achieve accurate semantic segmentation masks of medical images.
We evaluate the effectiveness of the proposed TransAttUnet on multiple medical image datasets from different imaging modalities. Our experiments mainly involve five typical challenges in clinical diagnosis, including: (1) Skin lesion segmentation on dermatoscopic images \cite{yu2018melanoma}; (2) Lung segmentation on chest X-ray images; (3) COVID-19 pneumonia lesion segmentation on chest CT images;  (4) Nuclei segmentation on divergent images \cite{rashno2017fully}, and (5) Gland segmentation on histology images  \cite{malik2020instance}.
Our main contributions are summarized as follows:
\begin{itemize}
\item This paper proposes a Transformer-Attention based U-shaped framework (TransAttUnet), that integrates the advantages of multi-level guided attention and multi-scale skip connections into the standard U-Net to improve segmentation performance for various medical images.

\item With the co-cooperation of the transformer self attention and global spatial attention, our TransAttUnet can achieve the strong ability to model contextual semantic information and global spatial relationships, which can guarantee the consistency of feature representations and semantic embeddings.

\item Compared with the one-step cascade connection, the proposed residual or dense step-growth connections not only can reduce the disturbance of noise, but also enable the model to mitigate the loss of fine details caused by directly upsampling with large scales.

\item Extensive experimental results on the five medical image datasets demonstrate the superiorities and generalizability of the proposed TransAttUnet for automatic medical image segmentation in comparison with state-of-the-art baselines.
\end{itemize}

The rest of this paper is organized as follows. Section~\ref{sec:2} reviews some related works of automatic medical image segmentation, and Section~\ref{sec:3} describes the proposed TransAttUnet. Next, the comprehensive experiments and visualization analyses are reported in Section~\ref{sec:4}. Finally, Section~\ref{sec:5} makes a conclusion of this work.

\section{Related Work}
\label{sec:2}
In this section, we make a brief overview of two related groups of works.
The first group mainly involves the state-of-the-art variants of U-Net for automatic medical image segmentation, while the second group introduces the applications of Transformer used in computer vision.

\subsection{Variants of U-Net }
Many efforts have been devoted to optimizing the structure of the U-Net in the field of automatic medical image segmentation. Typically, these variants can be roughly divided into two categories, i.e., attention guided approaches and multi-scale context approaches.

\subsubsection{\textbf{Attention Guided Approaches}}
Many attention-guided methods have been proposed to accurately segment the objects of interest in medical images of varying imaging modalities. To enhance the feature learning ability of U-Net, Attention U-Net \cite{oktay2018attention} utilizes the attention gate (AG)  to suppress irrelevant feature responses and highlight salient features, in order to improve the model sensitivity and prediction accuracy.
Channel-UNet\cite{chen2019channel} introduces the spatial channel convolution to determine the optimal mapping relationship among spatial information from different patches.
By simultaneously recalibrating the different types of features at the spatial and channel levels, SCAU-Net\cite{zhao2020scau} can guide the model to neglect irrelevant information and focus on more discriminant regions of the image. Similarly, 3D attention U-Net\cite{islam2019brain} applied the channel and spatial attention into the decoder subnetwork of 3D U-Net \cite{cciccek20163d} to segment brain tumors in MRI images.
XLSor\cite{tang2019xlsor} makes use of the criss-cross attention module to aggregate long-range pixel-wise contextual information in both horizontal and vertical directions for lung segmentation in chest X-ray images.
Residual Attention U-Net \cite{chen2020residual} applies the soft attention mechanism to improve the capability of the model to distinguish a variety of symptoms of the COVID-19 in chest CT images. However, the direct application of the attention mechanism might result in the loss of available feature representations, especially when the judgment for the region of interest goes wrong, which is inadequate to meet the needs of an ideal model.

\subsubsection{\textbf{Multi-Scale Context Approaches}}
To make full use of the multi-scale context information, prior works have attempted to combine the low-level features from the shallow layers with the high-level feature of the deep layers for retaining the detailed image information.
For example, UNet++\cite{zhou2018unet++} proposes a highly flexible multi-scale feature fusion scheme by aggregating features of varying semantic scales with redesigned skip connections. U$^{2}$-Net\cite{qin2020u2} exploits Residual U-blocks to capture intra-stage contextual information and directly cascades the inter-block feature mapping of the decoder subnetwork to mitigate the loss of fine details. MA-Unet\cite{cai2020ma} establishes a multi-scale mechanism to directly aggregating global contextual information of different scales from intermediate layers as final feature representations, and it also utilizes additional attention mechanisms to improve the prediction accuracy for medical image segmentation. However, one-step cascade connection may ignore some valuable details in the large-scale upsampling process, which still suffers from the challenge of information loss.

\subsection{Transformer in CV}
\subsubsection{\textbf{Transformer for Various Vision Tasks}}
With the development of Transformer used in various NLP tasks, more and more Transformer-based methods are developed for CV tasks. In particular, ViT\cite{dosovitskiy2020image} presents a pure self-attention Vision Transformer for image recognition, which is the first attempt of the transformer-based method to surpass the traditional CNN-based works.
By introducing Transformer into CNNs, DETR\cite{carion2020end} proposes a fully end-to-end object detector to eliminate the hand-designed components in CNN-based object detectors. Subsequently, SETR\cite{zheng2020rethinking} replaces the encoder in the standard FCN architecture with Transformer and achieves encouraging performance on the natural image segmentation task.

\subsubsection{\textbf{Transformer for Medical Image Segmentation}}
Inspired by ViT, TransUNet \cite{chen2021transunet} adopts Transformer as encoder and applies it with U-Net to enhance the performance of medical image segmentation tasks.
TransFuse \cite{zhang2021transfuse} combines Transformer and CNN in a parallel style to improve efficiency for modeling global context information. Similarly, MCTrans \cite{ji2021multi} utilizes Transformer to incorporate rich context modeling and semantic relationship mining for accurate biomedical image segmentation.
Besides, MedT\cite{valanarasu2021medical} proposes a Gated Axial-Attention model that utilizes Transformer based gated position-sensitive axial attention mechanism for medical image segmentation.
Unlike these works, the proposed TransAttUnet aims to investigate the feasibility of applying Transformer to overcome the inability of U-Net to model long-range contextual interactions.

\section{Methodology}
\label{sec:3}
This section mainly introduces the proposed TransAttUnet. Firstly, we give a brief overview of TransAttUnet. Then, we present the principles and the structure of TransAttUnet, followed by a detailed description of each component. Finally, the unified loss function used in our TransAttUnet is presented.

\subsection{Overview of TransAttUnet}
An input medical image $\bm{X} \in \mathbb{R}^{C\times H \times W}$, where $C$ is the number of channels and $H \times W$ represents the spatial resolution of image instance. The purpose of the automatic medical image segmentation task is to predict the corresponding pixel-wise semantic label maps with the size of $H \times W$. The general learning framework is illustrated in Fig. 2.
As with the previous works,  the proposed TransAttUnet is also built on the standard encoder-decoder U-shaped architecture, as shown in Fig.~\ref{fig:2}. To overcome the limitations mentioned in Section~\ref{sec:1}, TransAttUnet aims to leverage multi-level complementary self-aware attention components, as well as multi-scale skip connections \cite{zhou2018unet++}\cite{cai2020ma}  to further improve the semantic segmentation quality of medical images.
Compared with the standard U-Net, the SAA module in TransAttUnet can benefit greatly from both TSA and GSA mechanisms to capture the long-range contextual information, improving the representation ability of the encoder semantic features. Furthermore, the multi-scale skip connections used in TransAttUnet are designed to achieve the residual and dense shortcut connections between the intermediate layers of different semantic scales, which can aggregate the contextual information for multi-scale prediction fusion.

\subsection{Self-aware Attention Module}
Firstly, the proposed TransAttUnet augments the standard U-Net with a robust and effective self-aware attention module, that is positioned at the bottom of U-shaped architecture as the bridge between the encoder and decoder subnetworks. This duple mainly contains two independent self-attention mechanisms, i.e., transformer self attention (TSA) and global spatial attention (GSA),  which help capture the wider and richer contextual representations.

\subsubsection{\textbf{Transformer Self Attention}}
The TSA component is built on the multi-head self-attention function from Transformer, which allows the model to jointly attend to semantic information from global representation subspaces.
To capture the contextual information of absolute and relative position, the TSA component first introduces the learnt positional encoding to the input of the encoder features, which can be shared across all attention layers for a given query/key-value sequence.
The multi-head attention mechanism can be calculated separately in multiple single attention heads before being combined through another embedding.
The pipeline of transformer self attention component is depicted with the {\color{green}{green}} part in Fig.~\ref{fig:2}.

Specifically, the encoder features $\bm{F} \in \mathbb{R}^{c \times h \times w}$ is embedding into three inputs, including the matrix of queries $\bm{Q} \in \mathbb{R}^{c \times (h \times w)}$,  the matrix of keys $\bm{K} \in \mathbb{R}^{c \times (h \times w)}$, and $\bm{V} \in \mathbb{R}^{c \times (h \times w) }$.
\begin{equation}
\label{eq:0}
\begin{aligned}
\bm{Q} =  \bm{F}\cdot \bm{W}_q,
\bm{K} = \bm{F}\cdot \bm{W}_k,
\bm{V}= \bm{F}\cdot \bm{W}_v,
\end{aligned}
\end{equation}
where $\bm{W}_q$, $\bm{W}_k$, and $\bm{W}_v$ are the embedding matrices of different linear projections.
Then, a scaled dot-product operation with softmax normalization between $\bm{Q}$ and the transposed version of  $\bm{K}$ is conducted to generate the matrix of contextual attention map $\bm{A} \in \mathbb{R}^{c \times c}$, in which represents the similarities of given elements from $\bm{Q}$ with respect to global elements of $\bm{K}$. To obtain the aggregation of values weighted by attention weights, the contextual attention map $\bm{A}$ would be multiplied by $\bm{V}$. The multi-head attention can be formulated as:
\begin{equation}
\label{eq:1}
\begin{aligned}
TSA{(\bm{Q},\bm{K},\bm{V})}=softmax(\frac{\bm{Q}\bm{K}^\mathsf{T}}{\sqrt{d_k}}) \bm{V},
\end{aligned}
\end{equation}
where $\sqrt{d_k}$ is the dimensionality of query/key-value sequence.
Finally, we reshape the optimized feature maps to obtain the final output of TSA, i.e., $\bm{F}_{tsa} \in \mathbb{R}^{c\times h \times w }$.



\begin{figure*}
\centering
\subfigure[Cascade connection]{
\label{fig:6a}
\includegraphics[width=2.25in]{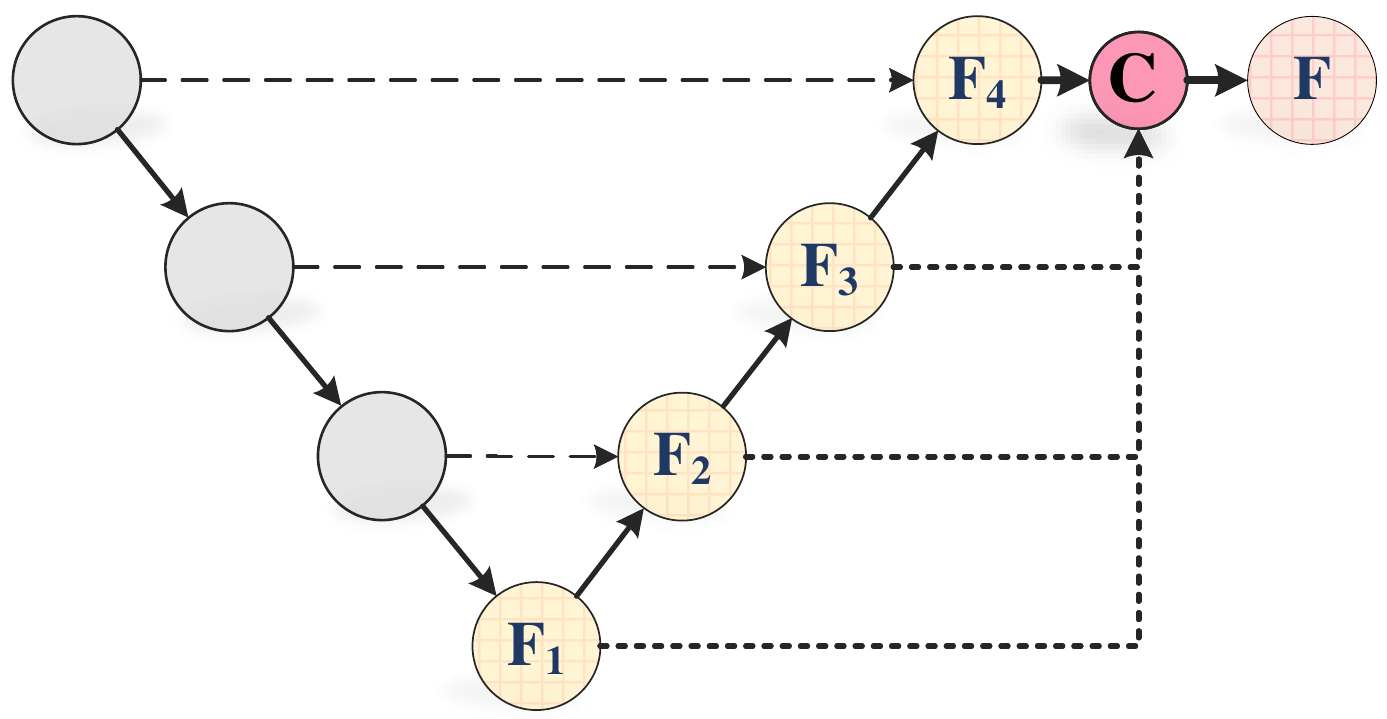} }
\subfigure[Residual connection]{
\label{fig:6b}
\includegraphics[width=2.25in]{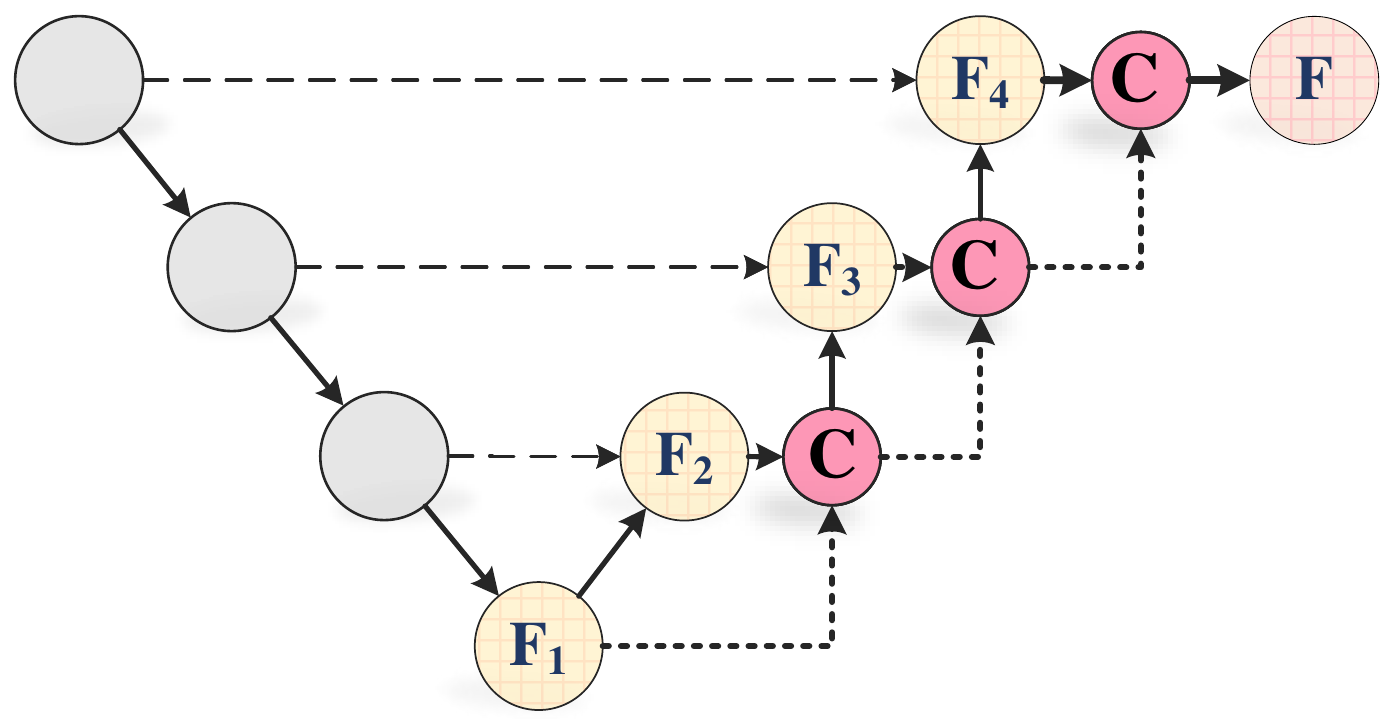} }
\subfigure[Dense connection]{
\label{fig:6c}
\includegraphics[width=2.25in]{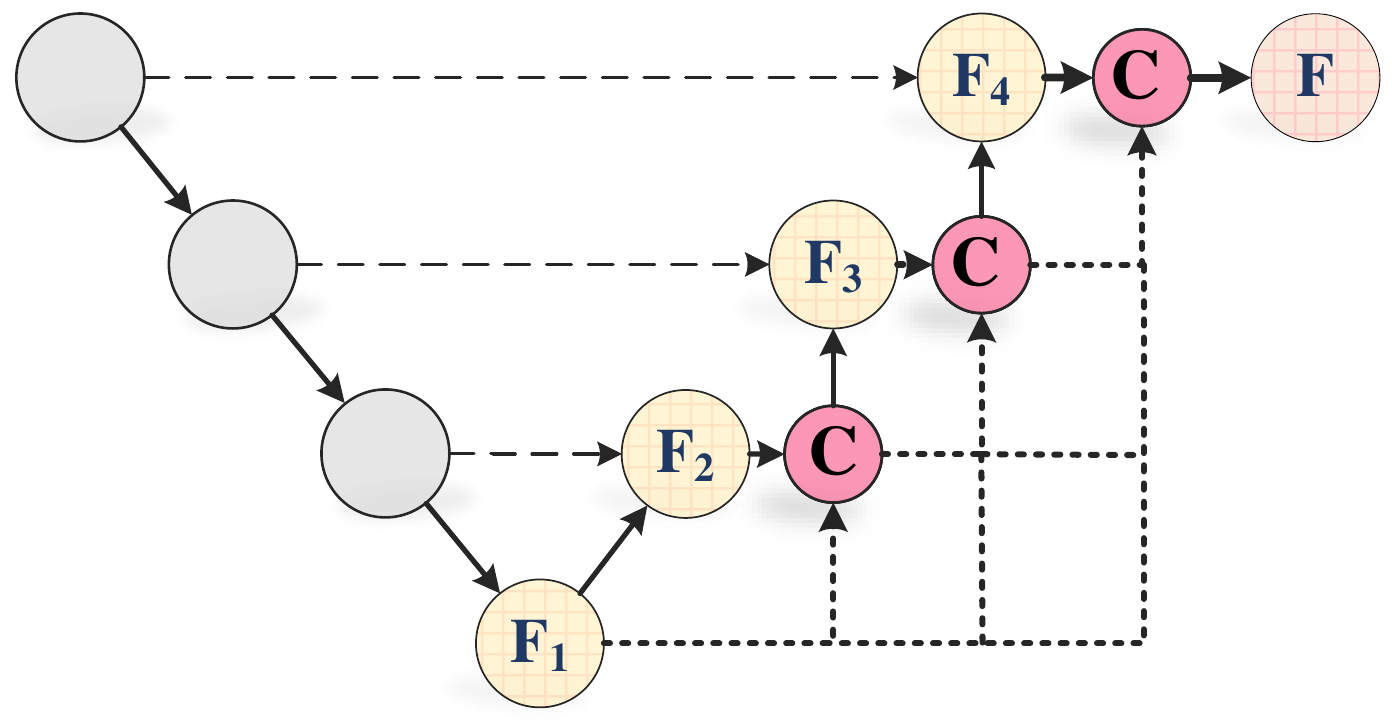} }
\caption{The comparison of the one-off cascade connections and our residual or dense step-growth connections.
(a) \textbf{Cascade connection}\cite{qin2020u2}\cite{cai2020ma}\cite{chen2021transunet}: The feature maps from all scales are directly concatenated to form a unified tensor;
(b) \textbf{Residual connection}: For each decoder block, the input and output features of the current block are concatenated and applied as the input of the subsequent block.
(c) \textbf{Dense connection}: For each decoder block, the upsampling features of previous decoder blocks are considered as inputs, and its own feature maps are used as inputs into all subsequent blocks.}
\label{fig:6}
\end{figure*}

\subsubsection{\textbf{Global Spatial Attention}}
Meanwhile, the SAA module also utilizes the GSA component to selectively aggregate global context to the learned features and encode broader contextual positional information into local features, which can improve the intra-class compact and optimize the feature representations. The architecture of global spatial attention is shown with the {\color{blue}{blue}} part in Fig.~\ref{fig:2}.

Firstly, two different types of convolutional operations are applied to the encoder features $\bm{F}_{en}$ to generate the feature maps $\bm{F}_{p}^{c'} \in \mathbb{R}^{ c' \times h \times w }$ and $\bm{F}_{p}^c \in \mathbb{R}^{c\times h \times w }$, $c'=c/8$.
Subsequently, $\bm{F}_{p}^{c'}$ is reshaped and transposed into feature maps $\bm{M} \in \mathbb{R}^{(h \times w) \times c' }$ and $\bm{N} \in \mathbb{R}^{c' \times (h \times w)}$, while $\bm{F}_{p}^c$ is transposed in to $\bm{W} \in \mathbb{R}^{c \times (h \times w)}$, respectively.
Then, a matrix multiplication operation with softmax normalization is performed on $\bm{M}$ and $\bm{N}$, resulting in the position attention maps  $\bm{B} \in \mathbb{R}^{(h \times w) \times (h \times w)}$, which can be defined as:
\begin{equation}
\label{eq:2}
\begin{aligned}
\bm{B}_{i, j}=\frac{exp(\bm{M}_i\cdot \bm{N}_j)}{\sum_{i=1}^{n}exp(\bm{M}_i\cdot \bm{N}_j)},
\end{aligned}
\end{equation}
where $\bm{B}_{i, j}$ measures the impact of $i^{th}$ position on $j^{th}$ position, and $n=h \times w$ is the number of pixels.
After that, $\bm{W}$ is multiplied with $\bm{B}$, and the resulting feature at each position can be formulated as:
\begin{equation}
\label{eq:3}
\begin{aligned}
GSA{(\bm{M},\bm{N},\bm{W})}_{p}=\sum_{q=1}^{h \times w}(\bm{W}_{q} \bm{B}_{p, q}).
\end{aligned}
\end{equation}
Similarly, we reshape the resulting features  to generate the final output of GSA, i.e., $\bm{F}_{gsa} \in \mathbb{R}^{c \times h \times w }$.

\subsubsection{\textbf{Attention Embedding Fusion}}
To make full use of the obtained contextual information and spatial relationships, a weighted combination scheme for the original and attention feature embeddings is used at the end of the SSA module, which is defined as:
\begin{equation}
\label{eq:4}
\begin{aligned}
F_{SAA}= \lambda_1 \bm{F}_{tsa} + \lambda_2 \bm{F}_{gsa} + \bm{F}_{en},
\end{aligned}
\end{equation}
where $\lambda_1$ and $\lambda_2$ are the scale parameters that controls the importance of the self attention maps and spatial attention maps, respectively. Both of them are initialized as 0 and are gradually increased to assign more weight to the important features.
In this way, we can further optimize the feature representations with semantic consistency.

\subsection{Multi-scale Skip Connection}
Notably, many advanced works \cite{qin2020u2}\cite{cai2020ma}\cite{chen2021transunet} have demonstrated the effectiveness of multi-scale feature fusion in encoding global and local contexts.  Specifically, the multi-scale skip connection scheme aims to aggregate the features of varying semantic scales with a series of transition operations, including upsampling, concatenation, and convolution. Inspired by previous works, three different types of connections, Cascade Connection, Residual Connection, and Dense Connection shown in Fig.~\ref{fig:6} are investigated in this study.

\subsubsection{\textbf{Cascade Connection}}  The feature maps of varying semantic scale from all the blocks are up-sampled to a common resolution through bilinear interpolation, and all of them would be directly concatenated into a unified feature representation, which is formulated as:
\begin{equation}
\label{eq:6}
\begin{aligned}
\bm{F}=f_n(\upsilon_1(\bm{F}_1) \oplus \upsilon_2(\bm{F}_2) \oplus …\oplus \bm{F}_n),
\end{aligned}
\end{equation}
where $\oplus$ denotes concatenation operations$, \upsilon_n({\cdot})$ and $f_n({\cdot})$ are the upsampling and mixed convolution operations in $n^{th}$ stage, respectively.

\subsubsection{\textbf{Residual Connection}} For each decoder block, the input feature maps are up-sampled to the resolution of outputs through bilinear interpolation, and then concatenated with the output feature maps as the inputs of the subsequent block, which is formulated as:
\begin{equation}
\label{eq:7}
\begin{aligned}
\bm{F}_{n}=f_{n}((\bm{F}_{n}) \oplus \upsilon_{n-1} (\bm{F}_{n-1})).
\end{aligned}
\end{equation}

\subsubsection{\textbf{Dense Connection}} The upsampling features of previous encoder blocks are integrated as the inputs of the current block, and the output feature maps are used as inputs into all subsequent blocks, which is formulated as:
\begin{equation}
\label{eq:8}
\begin{aligned}
\bm{F}_{n}=f_{n}(\upsilon_{1} (\bm{F}_{1}) \oplus \upsilon_{2} (\bm{F}_{2}) \oplus … \oplus \upsilon_{n-1}(\bm{F}_{n-1})).
\end{aligned}
\end{equation}

In particular, the proposed TransAttUnet focuses on two different multi-scale skip connection schemes, i.e., the residual connection and dense connection, to guide the upsampling process in the decoder subnetwork.
Compared to the existing works that only using the one-off cascade connection,  the residual or dense step-growth connections can gradually aggregate multiple decoder features of varying semantic scales to generate the most discriminative feature representations.
In this way,  the proposed TransAttUnet not only can mitigate the loss of fine details caused by over-upsampling, but also alleviate the problems of vanishing-gradient and overfitting.

\subsection{Loss Function}
In the training phase, the proposed TransAttUnet is trained with an objective function in an end-to-end manner.
The objective function is calculated by the Sorensen-Dice loss and Binary Cross-Entropy function with a pixel-wise soft-max over the final feature maps, which can be expressed as:
\begin{equation}
\label{eq:9}
\begin{aligned}
\centering
\begin{split}
\bm{\mathcal{L}}_{BCE} &=  \sum_{i=1}^{t}(y_i \log(p_i) + (1-y_i)\log(1-p_i )),\\
\bm{\mathcal{L}}_{Dice} &=1-\frac{\sum_{i=1}^{t}y_i p_i + \varepsilon}{\sum_{i=1}^{t}y_i +p_i + \varepsilon},\\
\bm{\mathcal{L}}&= \alpha\cdot \bm{\mathcal{L}}_{BCE}+ \beta\cdot \bm{\mathcal{L}}_{Dice},
\end{split}
\end{aligned}
\end{equation}
where $t$ is the total number of pixels in each image, $y_i $ represents the ground-truth value of the $i^{th}$ pixel, and $p_i$ is the confidence score of the $i^{th}$ pixel in prediction results. In our experiment, $\alpha = \beta=0.5$, and $\varepsilon =10^{-6}$.

\section{Experiments}
\label{sec:4}
In this section, we evaluate the performances of the proposed TransAttUnet framework on multiple benchmark datasets by comparing it with the state-of-the-art baselines. Next, we make a detailed discussion for the ablation studies. Finally, visualization analysis of decoder stages is presented.

\subsection{Datasets}
To verify the effectiveness and efficiency of our TransAttUnet, we first conducted comparative experiments for the task of skin lesion segmentation on ISIC-2018\cite{codella2019skin} dataset, and lung field segmentation on the combination of the JSRT\cite{shiraishi2000development}, Montgomery \cite{jaeger2014two}, and NIH \cite{tang2019xlsor} datasets.
Moreover, the proposed TransAttUnet is also evaluated on the Clean-CC-CCII dataset \cite{he2020benchmarking}, 2018 Data Science Bowl (Bowl) dataset\cite{caicedo2019nucleus}, and the Gland Segmentation (GlaS) dataset \cite{malik2020instance}.
Typically, the images from the CC-CCII, Bowl, and GLAS datasets might contain multiple segmenting objects with varying sizes and textures, which greatly increases the segmentation difficulty and complexity.

\subsubsection{\textbf{ISIC-2018}}
The ISIC-2018 is a large-scale dataset of dermoscopy images to develop the applications of automated diagnosis of melanoma from dermoscopic images.
It is provided for ISIC-2018 challenge\cite{codella2019skin} and contains three tasks, including lesion segmentation, lesion attribute detection, and disease classification. In particular, this paper focuses on the task of lesion segmentation from dermoscopic images by various types of dermoscopy.  It includes 2596 images with the corresponding annotations and these images are randomly split into 2076 images for training and 520 images for testing.

\subsubsection{\textbf{JSRT,  Montgomery \& NIH}}
Three datasets of frontal chest X-ray images, JSRT, Montgomery, and NIH are involved in our experiments for automatic lung field segmentation. The JSRT dataset comprises 247 chest X-ray images, among which 154 images are abnormal with pulmonary nodule and 93 images are normal. By contrast, the Montgomery dataset contains 138 chest X-ray images, including 80 normal patients and 58 patients with manifested tuberculosis. Different from the JSRT and Montgomery, the NIH dataset contains 178 chest X-ray images with various severity of lung diseases, which can hugely complicate the task of lung field segmentation.
Following the settings of previous works \cite{tang2019xlsor},  we combine these datasets and randomly split them into 407 images for training and 178 images for testing.

\subsubsection{\textbf{Clean-CC-CCII}}
As a publicly available chest CT dataset used in the field of automated COVID-19 diagnosis, the Clean-CC-CCII dataset contains thousands of annotated CT scans from 2,698 patients. In particular, it also provides researchers with high-quality annotation of infection marks to develop a robust model of  COVID-19  pneumonia lesion segmentation. In our experiment, 260 CT slices with pixel-wise annotations of COVID-19 pneumonia lesion are selected and randomly split into two subsets, i.e., a training set of 200 images and a test set of 60 images.

\subsubsection{\textbf{Bowl}}
 The Bowl dataset is established for the development of robust automatic nucleus segmentation algorithms.  It provides participants with a training set of 671 nuclei images along with pixel-wise masks for the nuclei and a test set of 3020 images, which are extracted from 15 diverse image sets of biological experiments. Note that each image contains dozens of nuclei with different sizes. Due to the lack of the annotation masks of the test set, we only evaluate the performance of the proposed method based on the training set. Following the settings of the existing work \cite{jha2020doubleu}, the training set is split into three subsets: 80\% for training, 10\% for validation, and 10\% for testing.

\subsubsection{\textbf{GlaS}}
The GlaS dataset is published by the Colon Histology Images Challenge Contest of MICCAl'2015 that aims to improve methods for quantifying the morphology of glands. It consists of 165 colon histology images derived from 16 H\&E stained histological sections of stage T3 or T4 colorectal adenocarcinoma from different patients. In particular, each sample is processed in the laboratory on different occasions, resulting in high inter-subject variability in both stain distribution and tissue structure. In our experiments, the GlaS dataset is split into two subsets: 85 images for training and 80 images for testing, which is consistent with the previous works \cite{valanarasu2020kiu}\cite{valanarasu2021medical}.

\subsection{Experimental Settings}
\subsubsection{\textbf{Baselines}}
In our experiments, three versions of TransAttUnet, i.e., TransAttUnet\_C, TransAttUnet\_D and TransAttUnet\_R are evaluated.
In particular, TransAttUnet\_C represents that the decoder blocks of TransAttUnet are linked with the one-off cascade connections\cite{qin2020u2}\cite{cai2020ma}\cite{chen2021transunet}. By contrast, TransAttUnet\_D utilizes the dense operations to connect the decoder blocks, while TransAttUnet\_R utilizes the residual operations to connect the decoder blocks.
In addition to the vanilla U-Net\cite{ronneberger2015u},  three broad approaches are involved in our comparative experiments as baselines, i.e., the attention-guided approaches, multi-scale context approaches, and Transformer-based approaches.
\begin{itemize}
\item \textbf{Attention-guided approaches}: Many advanced attention guided models,  i.e.,  Attention U-Net\cite{oktay2018attention}, Attention R2U-Net\cite{alom2018nuclei}, Channel-UNet\cite{chen2019channel}, XLSor\cite{tang2019xlsor}, and FANet\cite{tomar2021fanet}, and PraNet\cite{fan2020pranet}, are introduced to compare with  the proposed TransAttUnet.

\item \textbf{Multi-scale context approaches}: Meanwhile, several multi-scale context models are used as the major contenders, including Unet++\cite{zhou2018unet++}, R2U-Net\cite{alom2018nuclei}  ResUNet\cite{diakogiannis2020resunet}, ResUNet++\cite{jha2019resunet++}, BCDU-Net\cite{azad2019bi}, KiU-Net\cite{valanarasu2020kiu}, and DoubleU-Net\cite{jha2020doubleu}.

\item \textbf{Transformer-based approaches}: Moreover, some state-of-the-art Transformer-based models, i.e., MedT\cite{valanarasu2021medical}, MCTrans \cite{ji2021multi}, Swin-Unet\cite{cao2021swin}
and SegFormer\cite{xie2021segformer} are also considered as important baselines.
\end{itemize}

\subsubsection{\textbf{Implementation Details}}
To make a fair comparison with the existing works, the input images from Clean-CC-CCII, JSRT, Montgomery, and NIH are resized to $512\times 512$ for training and test, while the images of ISIC-2018 and Bowl are resized to $256\times 256$. Besides, the input images provided by GlaS are resized  to $128\times 128$ in a unified manner.
Note that, we employs 8 parallel attention heads in TSA module.
Moreover, we adopt the stochastic gradient descent (SGD) \cite{Loshchilov2017SGDRSG} optimizer with momentum $0.9$ and weight decay $0.0001$ to optimize the training process.
Furthermore, the proposed TransAttUnet is implemented by using the deep learning toolbox PyTorch \cite{Paszke2019PyTorchAI}, and all the experiments run on $1$ Nvidia Titan XP GPU with 12 GB memory.
The proposed TransAttUnet framework is trained for $100$ epochs with a batch size of $4$. Besides, the initial learning rate is $0.0001$, which decays by a factor of 10 for every $40$ epochs. In our experiments, the results are given as the probability maps directly outputted by the models, which can be binarized with a threshold of $0.5$ to get the binary masks for performance evaluation.

\subsubsection{\textbf{Evaluation Metrics}}
In our experiments, we adopt the mean Dice coefficient (DICE)\cite{anuar2010validate} as the key evaluation metric to measure the extent of similarity between the predicted mask and ground truth. Besides, four additional criteria, i.e., mean Intersection over Union (IoU), accuracy (ACC), recall (REC), and precision (PRE) scores are calculated pixel-wisely and used to evaluate the quantitative segmentation performance. These metrics are associated with four values, i.e., true-positive (TP), true-negative (TN), false-positive (FP), and false-negative (FN),

\begin{small}
\begin{equation}
\label{eq:8.1}
\begin{aligned}
{Dice}&=\displaystyle{\frac{2\times{TP}}{2\times{TP}+FP+FN}}, \\
{IoU}&=\displaystyle{\frac{{TP}}{{TP}+FP+FN}},\\
{Accuracy}&=\displaystyle{\frac{TP+TN}{TP+TN+FP+FN}}, \\
{Recall}&=\displaystyle{\frac{{TP}}{TP+FN}}, \\
{Precision}&=\displaystyle{\frac{{TP}}{TP+FP}}, \\
\end{aligned}
\end{equation}
\end{small}

\subsection{Experimental Results}
\subsubsection{\textbf{Evaluation on Skin Lesion Segmentation}}
\begin{figure*}
\centering
\subfigure[Quantitative results for skin lesion segmentation.]{
\label{fig:7a}
\includegraphics[width=7.05in]{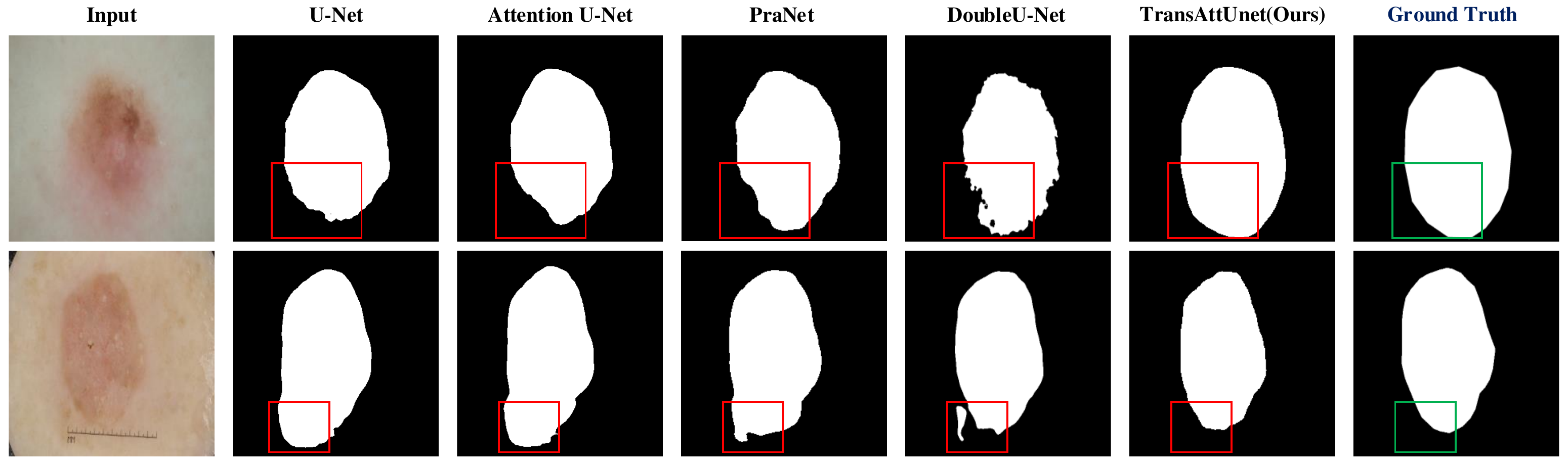} }
\subfigure[Quantitative results for lung field segmentation.]{
\label{fig:7b}
\includegraphics[width=7.05in]{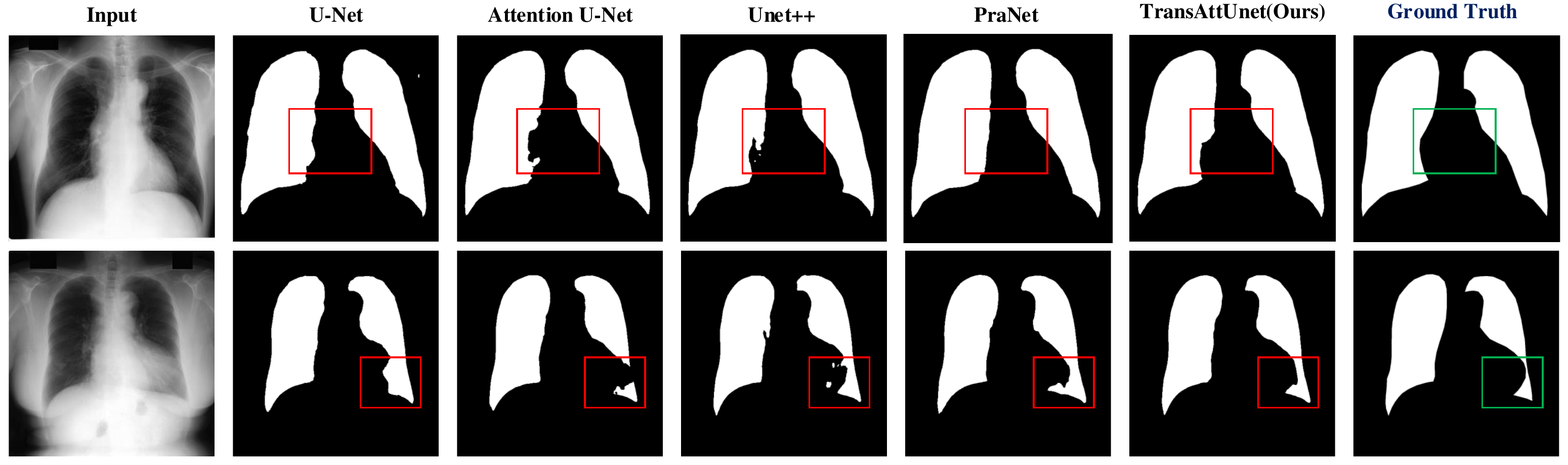} }
\subfigure[Quantitative results for pneumonia lesion segmentation.]{
\label{fig:7c}
\includegraphics[width=7.05in]{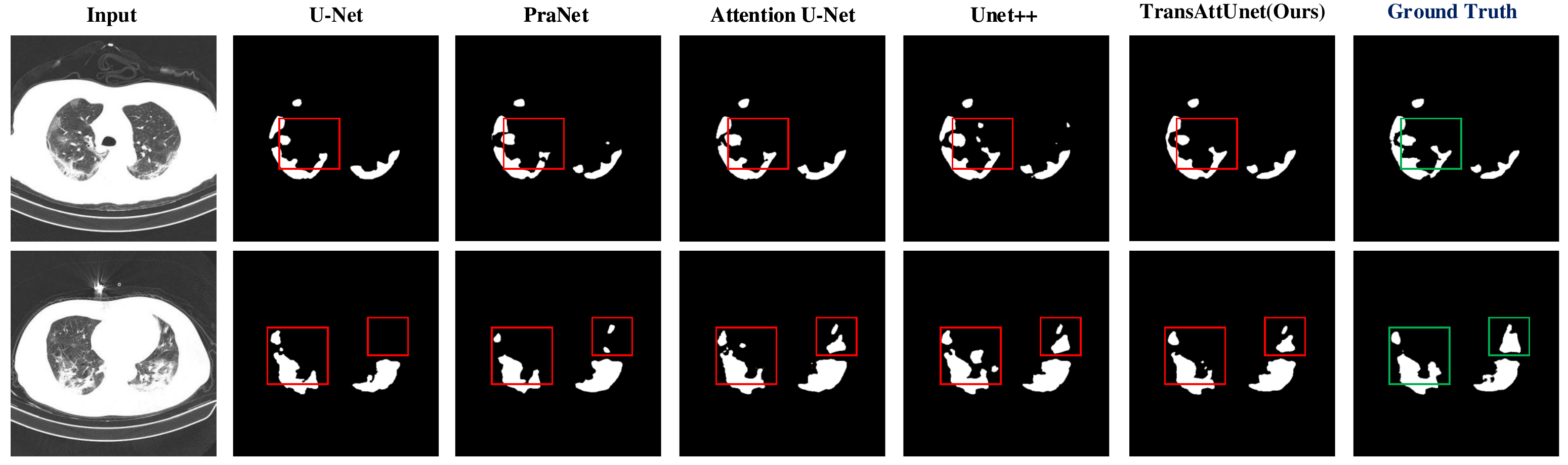} }
\caption{Comparison of quantitative results between the state-of-the-art baselines and the proposed TransAttUnet on the (a) ISIC 2018 dataset,  (b) the combination of JSRT, Montgomery, and NIH datasets, (c) the Clean-CC-CCII dataset, respectively.
To make better visualize the differences between lung segmentation results and ground truths, we highlight the key region with the appropriate boxes.}
\label{fig:7}
\end{figure*}

To evaluate the effectiveness of the proposed TransAttUnet, we first conduct the experiments on the ISIC-2018 dataset for the task of skin lesion segmentation.
The comparison results of evaluation metrics are presented in TABLE~\ref{tab:1}, and the corresponding quantitative results are illustrated in Fig.~\ref{fig:7a}.

From Table~\ref{tab:1}, we have the following observations:
1) Comparing with the vanilla U-Net (67.40\%), we can observe that the attention-guided models, such as Channel-UNet (84.82\%) and PraNet (87.46\%), are obviously superior to the vanilla U-Net (67.40\%) with the guidance of various attentional mechanisms. These improvements can demonstrate that attention-aware mechanism can play a vital role in medical image segmentation.
2) By aggregating the context information from varying semantic scales, it is easy to see that the multi-scale context approaches, i.e., BCDU-Net  (85.10\%) and DoubleU-Net (89.62\%), have the powerful ability in dealing with the task of skin lesion segmentation, which proves the effectiveness of the multi-scale context fusion scheme.
3) By contrast, the recent Transformer-based framework, i.e, MCTrans (90.35\%), is superior to the above works. It can demonstrate the superiority of the Transformer in learning the long-range dependencies of different pixels.
4) Although MCTrans can achieve reliable performance, the proposed TransAttUnet still outperforms MCTrans (90.74\% vs. 90.35\%), which verifies the efficacy of jointly exploring multi-level guided attention and multi-scale skip connection. Moreover, our TransAttUnet achieves the highest scores on almost all evaluation metrics and shows a more precise and fine segmentation output of the proposed network than the existing baselines, as shown in Fig.~\ref{fig:7a}.
5) Both the diagnostic sensitivity of the proposed TransAttUnet\_R and TransAttUnet\_D outperform TransAttUnet\_C, which demonstates that the residual or dense step-growth connections can better aggregate multiple decoder features of varying semantic scales rather than the one-off cascade connections. Furthermore, the proposed TransAttUnet\_R surpasses the DICE score of TransAttUnet\_D by 0.6\%. That is probably because the dense connection used in TransAttUnet might lead to a fair amount of redundant information.
Therefore, these comparative results demonstrate the effectiveness of the proposed TransAttUnet for automated skin lesion segmentation.
\begin{table}[t]
\centering
\caption{Comparisons with the state-of-the-art baselines on the ISIC-2018 dataset. All results were analysed in percentage (\%) terms. Results of the model with ``*" are reimplemented by the released source codes. The  ``-" denotes the corresponding result is not provided. For each column, the best and second best results are highlighted in {\color{red}{red}} and {\color{blue}{blue}}, respectively.}
\label{tab:1}
\renewcommand\tabcolsep{5.5pt}
\renewcommand\arraystretch{1.5}
\begin{tabular}{l|c|ccccc}
\toprule
\textbf{Method} & Year     & DICE  & IoU    & ACC   & REC  & PRE \\
\midrule
U-Net\cite{ronneberger2015u}       &    2015     & 67.40          & 54.90      &-    & 70.80          & -                  \\
Attention U-Net\cite{oktay2018attention}  & 2018   & 66.50          & 56.60   &-       & 71.70          & -                  \\
R2U-Net\cite{alom2018nuclei}        &  2018    & 67.90          & 58.10      &-    & 79.20          & -                  \\
Att R2UNet\cite{alom2018nuclei}  &2018  & 69.10          & 59.20   &-       & 72.60          & -                  \\
ResUNet*\cite{jha2019resunet++} &2019   & 79.15          & 70.15   & 92.28      & 82.43          & 84.77                  \\
Channel-UNet*\cite{chen2019channel} &2019& 84.82  &  75.92    & 94.10  & 94.01  &  81.04 \\
BCDU-Net\cite{azad2019bi}   & 2019   & 85.10          & -        &-       & 7850          & -                  \\
FANet\cite{tomar2021fanet}    &  2021    & 87.31          & 80.23     &-     & 86.50          & 92.35             \\
PraNet*\cite{fan2020pranet}   & 2021& 87.46          & 80.23     &  95.37   &  \color{red}{91.28}          & 87.59             \\
DoubleU-Net\cite{jha2020doubleu}  &2020  & 89.62          & 82.12     &-     & 87.80          & \color{red}{94.59}    \\
Swin-Unet*\cite{cao2021swin} & 2021  &  89.72    &  82.90    & -   & 90.32  & 92.04    \\
SegFormer*\cite{xie2021segformer} & 2021  &  90.24    &  83.60   & -   & 91.12  & 92.10    \\
MCTrans \cite{ji2021multi} &2021  & \color{blue}{90.35}          & -     &-     & -          & -    \\
\midrule
TransAttUnet\_C &-&   89.25  & 81.46  &  95.06  &    89.90    &  91.59  \\
TransAttUnet\_D & -& 90.14          & \color{blue}{83.04}     & \color{blue}{96.14}    & 90.42          & 92.17             \\
TransAttUnet\_R &- & \color{red}{90.74} &\color{red}{83.80} & \color{red}{96.38} &\color{blue}{90.93} & \color{blue}{92.42}             \\
\bottomrule
\end{tabular}
\end{table}

\begin{table}[h]
\centering
\caption{Comparisons with the state-of-the-art baselines on the JSRT and Montgomery dataset. All results were analysed in percentage (\%) terms. Results of the model with ``*" are reimplemented by the released source codes. The  ``-" denotes the corresponding result is not provided. For each column, the best and second best results are highlighted in {\color{red}{red}} and {\color{blue}{blue}}, respectively.}
\label{tab:2}
\renewcommand\tabcolsep{5.5pt}
\renewcommand\arraystretch{1.5}
\begin{tabular}{l|c|ccccc}
\toprule
\textbf{Method}  & Year    & DICE  & IoU    & ACC   & REC  & PRE \\
\midrule
U-Net\cite{ronneberger2015u}      & 2015         & 96.17          & 92.71      &98.21    & 94.94          & 97.50                 \\
XLSor\cite{tang2019xlsor}   &2019      & 97.54    & -      &-    & 97.40          & 97.73                 \\
ResUNet*\cite{diakogiannis2020resunet}  &2020   & 97.12          &94.45       &98.64       & 96.61          & 97.70                  \\
Attention U-Net* \cite{oktay2018attention} &2018  & 97.59          &95.31       &98.81       & 98.82          & 96.41                  \\
Swin-Unet*\cite{cao2021swin}  & 2021 & 97.67 & 95.48 & 98.71 & 95.42 & 98.36  \\
Unet++*\cite{zhou2018unet++} & 2018  & 97.84        & 95.80   & 98.93      & 99.28          & 96.47                  \\
ResUNet++*\cite{jha2019resunet++}  &2019     & 97.92          & 95.95   &98.96       & 98.68   & 98.48     \\
FANet*\cite{tomar2021fanet}&2021     & 98.28  &  96.64    & 99.12  & 98.04  &  \color{blue}{98.54} \\
PraNet*\cite{fan2020pranet}  &2021   & 98.36          & 96.80     &  99.17   & 98.26          & 98.48             \\
\midrule
TransAttUnet\_C &-&    98.14    &  96.37 &  99.08 &  98.56 & 98.15 \\
TransAttUnet\_D &-& \color{blue}{98.56}       &  \color{blue}{97.18}     &  \color{blue}{99.27}     &\color{red}{98.88}       & 98.26            \\
TransAttUnet\_R &-& \color{red}{98.88} &\color{red}{97.82} &\color{red}{99.41} & \color{blue}{98.74} &\color{red}{99.04}             \\
\bottomrule
\end{tabular}
\end{table}

\begin{table}
\centering
\caption{Comparisons with the state-of-the-art baselines on the Clean-CC-CCII dataset. All results were analysed in percentage (\%) terms. Results of the model with ``*" are reimplemented by the released source codes. The  ``-" denotes the corresponding result is not provided. For each column, the best and second best results are highlighted in {\color{red}{red}} and {\color{blue}{blue}}, respectively.}
\label{tab:3}
\renewcommand\tabcolsep{5.5pt}
\renewcommand\arraystretch{1.5}
\begin{tabular}{l|c|ccccc}
\toprule
\textbf{Method}    &Year    & DICE  & IoU    & ACC   & REC  & PRE \\
\midrule
U-Net*\cite{ronneberger2015u}       &2015         & 82.39          & 71.25      &99.19    & 78.99          & 87.62                  \\
PraNet*\cite{tomar2021fanet}   &2021  & 82.40          & 71.26     &  99.14   & 82.53          & 83.74             \\
ResUNet*\cite{jha2019resunet++}  &2019   & 82.90          & 71.96   & 99.17      & 83.60          & 83.35                  \\
Attention U-Net* \cite{oktay2018attention}   &2018   & 83.92          & 73.40   &99.24       & 83.23          & 86.07         \\
Swin-Unet\cite{cao2021swin} & 2021 & 84.47 & 76.59 & 94.95 & 84.60 & 84.35  \\
Unet++*\cite{zhou2018unet++}  &2018  & 84.64        & 74.43   & 99.30      & 81.98          & \color{blue}{86.52}                 \\
PraNet*\cite{fan2020pranet} &2021 & 84.82 & 76.26 & 93.91 & 85.60 & 85.08 \\
SegFormer\cite{xie2021segformer} & 2021  & 84.96 & 76.56 & 94.31 & 85.76 & 85.63  \\
ResUNet++*\cite{jha2019resunet++}  &2019   &   85.17        & 75.36   & 99.27      & \color{blue}{86.40}          & 85.61                  \\
\midrule
TransAttUnet\_C &-&   85.51 & 75.56 &  99.23  & 83.54 & 85.52  \\
TransAttUnet\_D &-& \color{blue}{86.08}          & \color{blue}{76.25}     & \color{blue}{99.32}    &  \color{red}{86.91}          & 86.23          \\
TransAttUnet\_R &-&\color{red}{86.57} &\color{red}{77.16} & \color{red}{99.38} &85.95 &\color{red}{88.47}             \\
\bottomrule
\end{tabular}
\end{table}

\subsubsection{\textbf{Evaluation on Lung Field Segmentation}}
Moreover, we evaluate our TransAttUnet to solve the task of lung field segmentation from the chest X-ray images on the combination of JSRT,  Montgomery, and NIH datasets. The comparison results of evaluation metrics are presented in TABLE~\ref{tab:2} and the corresponding quantitative results are illustrated in Fig.~\ref{fig:7b}.

Based on the experimental results in Table~\ref{tab:2}, we have some new observations as follows:
1) Firstly, we can observe that the corresponding scores of the evaluation metrics are quite close. That is because that the testing chest X-ray images provided by the JSRT and Montgomery datasets are generally normal, and the noise interference caused by various lesion features is reduced in semantic segmentation.
2) Nevertheless, the proposed TransAttUnet achieves better evaluation results than previous state-of-the-art models and yields the highest DICE score of 98.88\% for lung field segmentation, which demonstrates the efficacy of the designed TransAttUnet. As shown in Fig.~\ref{fig:7b},  the segmentation outputs of our TransAttUnet are more close to the ground truths in comparison with other baselines.
3) In particular, the performance of TransAttUnet is clearly superior to the previous baselines, especially for U-Net (96.17\%) with the improvement of 2.71\% in terms of the DICE score.
This improvement demonstrates that the proposed TransAttUnet benefits greatly from encoder-decoder guided attention and multi-scale skip connections, which help learn the global contextual information and discriminative features to distinguish the lung field from the surrounding structures.
4) Likewise, the proposed TransAttUnet consistently outperforms the recent works, i.e., FANet (98.28\%) and PraNet (98.36\%), which can verify the abilities of TransAttUnet in improving the segmentation quality of the details.

\subsubsection{\textbf{Evaluation on Pneumonia Lesion Segmentation}}
Due to the epidemic of COVID-19 in recent years, it has been a hot topic in the field of medical image analysis. Note that it is difficult to accurately and completely identify the segmenting objects of pneumonia lesions from chest CT due to the complex textures and shapes. Thus, we also perform the experiments on the Clean-CC-CCII dataset to further ensure that the proposed TransAttUnet is appropriate for pneumonia lesion segmentation. The comparison results of evaluation metrics are presented in TABLE~\ref{tab:3} and the corresponding quantitative results are illustrated in Fig.~\ref{fig:7c}.

On the basis of the above results, we have the following observations:
1)  It can be observed that the proposed TransAttUnet consistently outperforms the previous baselines and yields the highest DICE score of 86.57\%, which again demonstrates the superiorities of our TransAttUnet method.
2) Both TransAttUnet\_D and TransAttUnet\_R are able to obtain better segmentation performance than TransAttUnet\_C and the previous baselines. By contrast, the proposed TransAttUnet\_R achieves a better performance than TransAttUnet\_D (86.57\%  vs. 86.08\%), which further verifies the superiorities of the residual connection in comparison with the dense connection.
3) However, some reliable models, such as PraNet, achieve the relatively low Dice score in the COVID-19 pneumonia lesion segmentation task. We suspected that it might suffer from the overfitting problem due to the lack of training data.
4) Besides, the proposed TransAttUnet outperforms the recent Transformer-based work, i.e., Swin-Unet (84.47\%) and SegFormer (84.96\%), which demonstrates the powerful ability of TransAttUnet in medical image segmentation.
5) Futhermore, Fig.~\ref{fig:7c} shows that the superiority of our TransAttUnet over the other methods when dealing with lesions at different scales, which proves the availability of the proposed TransAttUnet for Covid-19 pneumonia lesion segmentation.

\begin{table}[t]
\centering
\caption{Comparisons with the state-of-the-art baselines on the 2018 Data Science Bowl dataset. All results were analysed in percentage (\%) terms. Results of the model with ``*" are reimplemented by the released source codes. The  ``-" denotes the corresponding result is not provided. For each column, the best and second best results are highlighted in {\color{red}{red}} and {\color{blue}{blue}}, respectively.}
\label{tab:4}
\renewcommand\tabcolsep{5.5pt}
\renewcommand\arraystretch{1.5}
\begin{tabular}{l|c|ccccc}
\toprule
\textbf{Method}   & Year   & DICE  & IoU    & ACC   & REC  & PRE \\
\midrule
U-Net\cite{ronneberger2015u}      &  2015        & 75.73          & 91.03      &-    & -          & -                 \\
FANet*\cite{tomar2021fanet}  & 2021  & 81.03          & 71.08     &  95.59   & 80.62          & 82.31             \\
Channel-UNet*\cite{chen2019channel}&2019 & 87.55  &  79.75    & 96.27  & 90.70  &  87.86 \\
Unet++\cite{zhou2018unet++}  & 2018 & 89.74        & \color{red}{92.55}   & -      & -          & -                  \\
ResUNet*\cite{diakogiannis2020resunet}  & 2020  & 89.91          &82.44       &97.05       & 90.00          & 90.84                  \\
Attention U-Net \cite{oktay2018attention}  &2018 & 90.83          &91.03       &-       & -          & 91.61                  \\
PraNet*\cite{fan2020pranet}  & 2021 & 90.85 & 85.34 & - & - & 90.94\\
DoubleU-Net\cite{jha2020doubleu}  & 2020 & 91.33 &84.07 &- &64.07 &\color{red}{94.96}    \\
\midrule
TransAttUnet\_C &-&    90.04    & 84.36  &  97.05 &    90.03    &  91.23  \\
TransAttUnet\_D &- &\color{blue}{91.34}       & 84.62     & \color{blue}{97.37}     &\color{red}{91.86}       & 91.53             \\
TransAttUnet\_R &- & \color{red}{91.62} & \color{blue}{84.98} & \color{red}{97.46} &\color{blue}{91.85} &\color{blue}{91.93}             \\
\bottomrule
\end{tabular}
\end{table}

\begin{table}[t]
\centering
\caption{Comparisons with the state-of-the-art baselines on the Glas dataset. All results were analysed in percentage (\%) terms. Results of the model with ``*" are reimplemented by the released source codes. The  ``-" denotes the corresponding result is not provided. For each column, the best and second best results are highlighted in {\color{red}{red}} and {\color{blue}{blue}}, respectively.}
\label{tab:5}
\renewcommand\tabcolsep{5.5pt}
\renewcommand\arraystretch{1.5}
\begin{tabular}{l|c|ccccc}
\toprule
\textbf{Method}  & Year    & DICE  & IoU    & ACC   & REC  & PRE \\
\midrule
U-Net\cite{ronneberger2015u}  &2015 & 75.73          & 91.03      &-    & -          & -                 \\
ResUNet*\cite{diakogiannis2020resunet} &  2020  &80.88          &69.11       &81.49       & 85.11          & 80.01           \\
MedT\cite{valanarasu2021medical}&2021 & 81.82         & 69.61      &-    & -          & -                 \\
Unet++\cite{zhou2018unet++}  & 2018 & 81.83        & 69.61   & -      & -          & -                  \\
Attention U-Net \cite{oktay2018attention} & 2018 & 81.59          &70.06       &-       & -          & -                 \\
KiU-Net\cite{valanarasu2020kiu}  & 2020 & 83.25 &72.78 &- &- &-    \\
FANet\cite{tomar2021fanet}  & 2021 & 84.67 & 74.30 & - & - & -   \\
Swin-Unet\cite{cao2021swin} & 2021 & 86.70 & 77.32 & - & 89.00 & 86.12   \\
SegFormer\cite{xie2021segformer}  & 2021  &  87.36    &  79.71   &  -  & 85.56  &  86.53  \\
\midrule
TransAttUnet\_C &-&  87.35 & 79.55  & 95.29 & 87.82  &  86.46 \\
TransAttUnet\_D &-&\color{blue} 88.37       & \color{blue}80.08     & \color{blue}88.47     &\color{blue}89.19       & \color{blue}88.49             \\
TransAttUnet\_R &-&\color{red}{89.11} & \color{red}{81.13} & \color{red}{89.02} &\color{red}{90.08} &\color{red}{88.95}             \\
\bottomrule
\end{tabular}
\end{table}

\begin{figure*}[t]
\centering
\subfigure[Quantitative results for nuclei segmentation.]{
\label{fig:8a}
\includegraphics[width=7.05in]{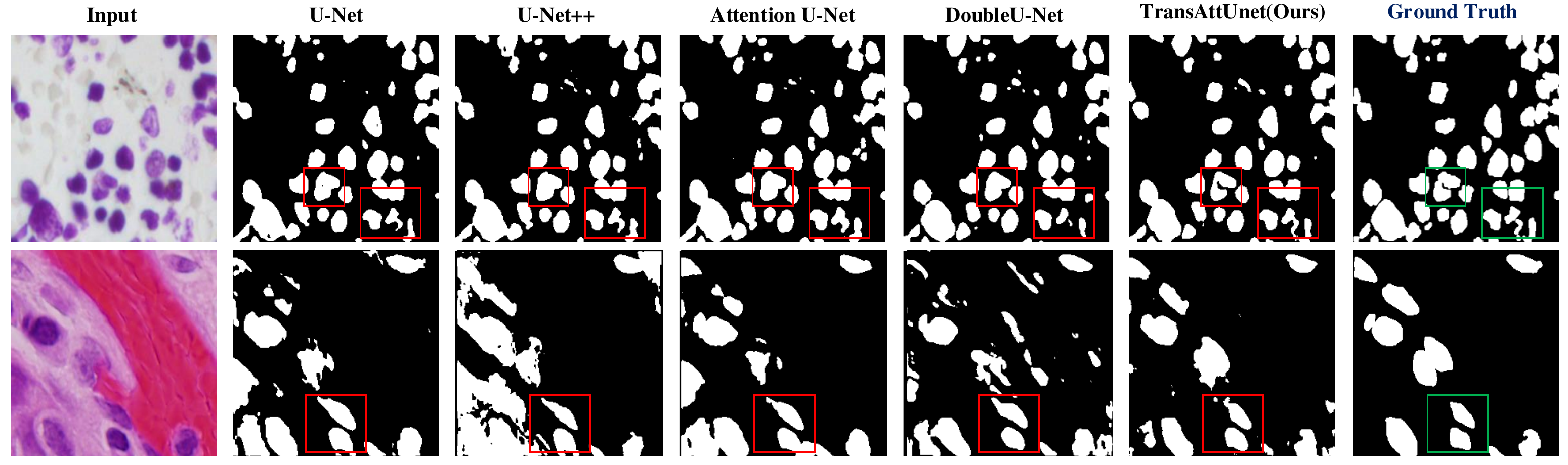} }
\subfigure[Quantitative results for gland segmentation.]{
\label{fig:8b}
\includegraphics[width=7.05in]{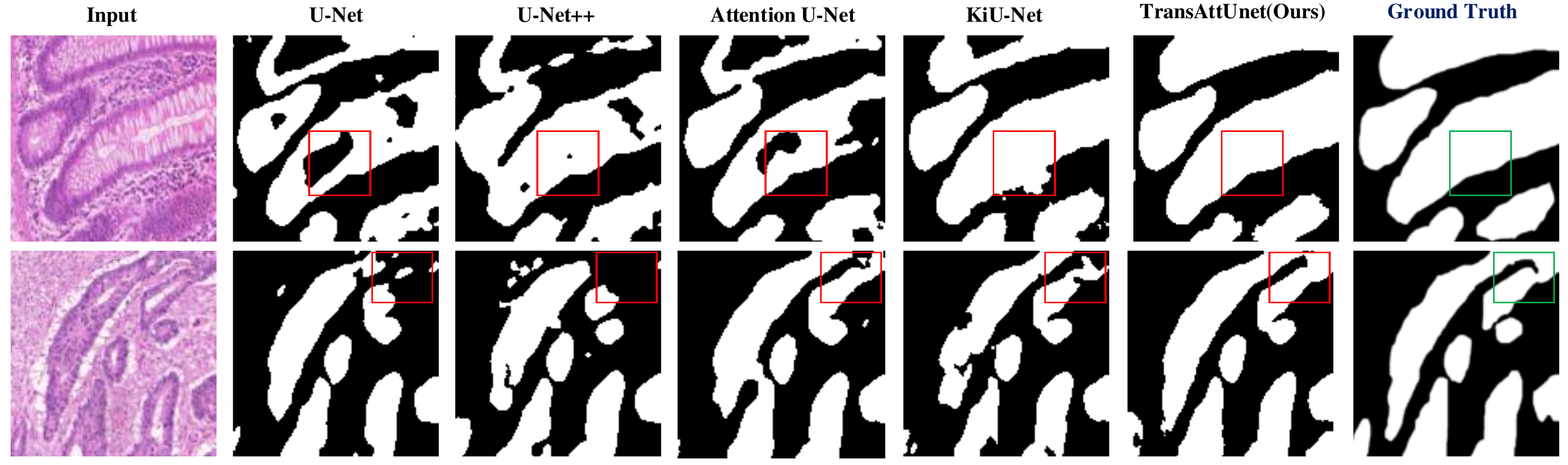} }
\caption{Comparison of quantitative results between the state-of-the-art baselines and the proposed TransAttUnet on the (a) 2018 Data Science Bowl dataset, (b) the GLAS dataset, respectively.
To make better visualize the differences between lung segmentation results and ground truths, we highlight the key region with the appropriate boxes.}
\label{fig:8}
\end{figure*}

\subsubsection{\textbf{Evaluation on Nuclei  Segmentation}}
In this part, we evaluate the proposed TransAttUnet on the 2018 Data Science Bowl dataset for multiple nuclei segmentation.
The comparison results of evaluation metrics are presented in TABLE~\ref{tab:4} and the corresponding quantitative results are illustrated in Fig.~\ref{fig:8a}.

Based on the above comparative results, we have the following observations:
1)  Compared with the vanilla U-Net, the use of encoder-decoder guided attention and multi-scale skip connections can lead to the further improvement of segmentation quality, improving with the DICE score from 75.76\% to 91.92\%.
2)  Consistently,  the proposed TransAttUnet is superior to the existing competitors of the attention-guided and multi-scale context approaches, such as Attention U-Net (91.62\% vs. 90.93\%) and ResUNet (91.62\% vs. 89.91\%).
3) It can be seen that our TransAttUnet yields the highest score on almost all evaluation metrics. Although DoubleU-Net outperforms in terms of precision, our TransAttUnet produces the higher scores on other metrics, especially for the IoU score of 84.98\% with the improvement of 0.91\%.
4) As illustrated in Fig.~\ref{fig:8a}, we can observe that our TransAttUnet can effectively capture the boundaries of cell nuclei and generate better segmentation prediction. These comparative results can prove the efficacy of the proposed method for identifying multiple segmenting objects.

\begin{figure*}[t]
\centering
\includegraphics[width=7.1in]{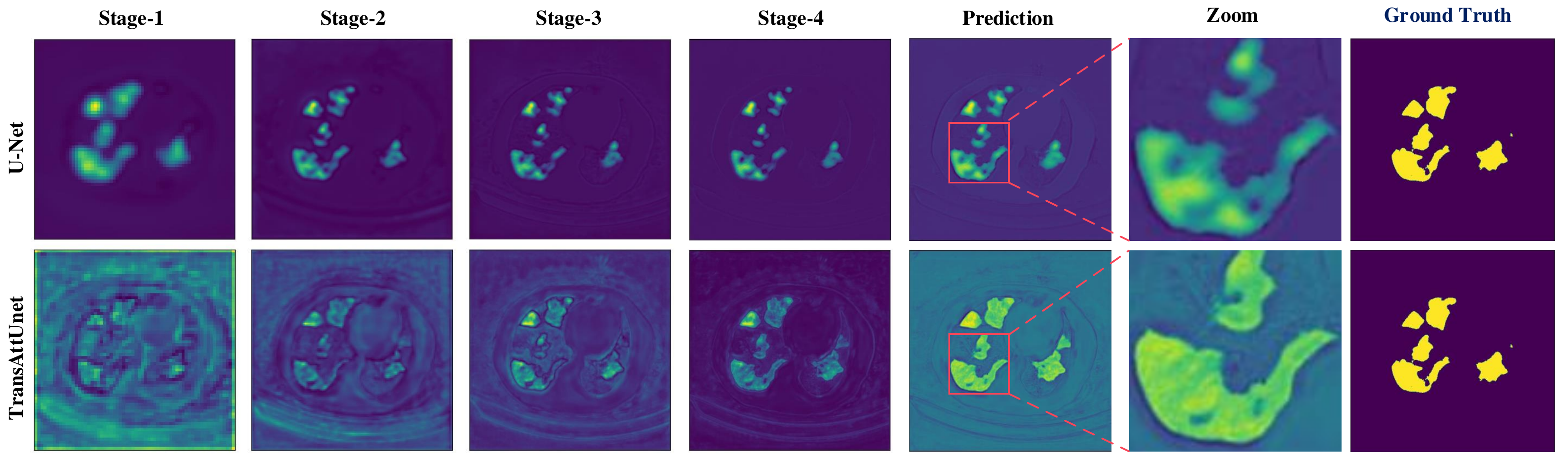}
\caption{Visualizations of feature maps produced by vanilla U-Net and the proposed TransAttUnet in different decoder stages based on the Clean-CC-CCII dataset. Best viewed with zoom in.}
\label{fig:9}
\end{figure*}

\subsubsection{\textbf{Evaluation on Gland Segmentation}}
Furthermore, we also conduct comparative experiments on the GLAS dataset to demonstrate the validity of the proposed TransAttUnet for quantifying the morphology of glands.
The comparison results of evaluation metrics are presented in TABLE~\ref{tab:5} and the corresponding quantitative results are illustrated in Fig.~\ref{fig:8b}.

According to these experimental results, we have the following observations:
1) Note that GLAS is a small dataset that contains multiple complex objects of interest. By incorporating with the encoder-decoder guided attention and multi-scale skip connections, our TransAttUnet achieves a better performance than the existing baselines and contributes a new state-of-the-art technique for automatic gland segmentation.
2) In particular, the proposed TransAttUnet outperforms the recent Transformer-based work, i.e.,  MedT  (81.82\%), Swin-Unet (86.70\%) and SegFormer (87.36\%), which again demonstrates the powerful ability of TransAttUnet in medical image segmentation.
3) Moreover, we can clearly see that both TransAttUnet\_D and  TransAttUnet\_R clearly outperform TransAttUnet\_C and the previous state-of-the-art method, i.e., KiU-Net. In particular, TransAttUnet\_D achieves the improvement of 5.86\% and 8.85\%  in terms of DICE and IoU scores, respectively. According to this significant improvement, it has been proven the reliability and superiority of our TransAttUnet.
4) Besides, Fig.~\ref{fig:8b} demonstrates that our TransAttUnet can better distinguish the gland itself from the surrounding tissue, leading to excellent gland segmentation performance.

All results in the above experiments quantitatively demonstrate the effectiveness and generalizability of the proposed TransAttUnet for medical image segmentation in various challenging scenarios.

\subsection{Ablation Studies}
To evaluate the effectiveness of each component added in the proposed TransAttUnet, we conduct comprehensive ablation experiments by removing the components successively. The experimental results are presented in Table~\ref{tab:6}. In detail,  ``TSA" denotes the proposed transformer self attention block; ``GSA" denotes the designed global spatial attention block; ``MSC" denotes the multi-scale skip connections between the decoders; and ``TAU" is considered as the ``full" model.

\begin{table}[t]
  \caption{Comparisons of mean DICE score for ablation studies with different experimental settings. For each column, the best results are highlighted in {\color{red}{red}}.}
  \label{tab:6}
  \renewcommand\tabcolsep{7.0pt}
  \renewcommand{\arraystretch}{1.3}
  \centering
  \begin{tabular}{l|ccccc}
    \toprule
    Methods &Skin   &  Lung  &  Pneu. & Bowl & Gland  \\
    \midrule
    TAU w/o TSA + GSA &  83.89    & 97.55 &  83.56  & 84.54   &81.59  \\
    TAU w/o TSA + MSC & 84.49    & 97.75  &  84.99  &  85.82  &83.84  \\
    TAU w/o GSA + MSC &  85.12   & 97.79 &  85.03 & 86.55   &84.64  \\
    TAU w/o TSA           &  87.13   &  98.16 &  85.59  & 88.42   &85.96 \\
    TAU w/o GSA           & 87.86    &  98.24 & 85.63 & 88.91   &86.78  \\
    TAU w/o MSC            &  88.55    & 98.49 &  85.97 &90.37    &87.83  \\
    \midrule
    TransAttUnet&  \color{red}{90.74} &  \color{red}{98.88} &   \color{red}{86.57} &  \color{red}{91.62} &  \color{red}{89.11}\\
    \bottomrule
  \end{tabular}
\end{table}

When both TSA and GSA blocks are removed, ``TAU w/o TSA + GSA" would suffer serious performance degradation. Nonetheless,  it consistently outperforms the vanilla U-Net, which demonstrates the effectiveness of the multi-scale skip connections between the decoders. Moreover,  the results of ``TAU w/o TSA + MSC" and ``TAU w/o GSA + MSC" can indicate that the proposed TSA and GSA blocks are equally effective in developing the segmentation quality.
Subsequently, we also perform additional ablation studies by removing multiple components to further verify the effectiveness of our work.
For example, after removing the component of MSC, the evaluation scores of ``TAU w/o MSC" drop to 88.55\%, 98.49\%, 85.97\%, 90.37\%, and 87.83\%, respectively. Likewise, the performance of ``TAU w/o TSA" and ``TAU w/o GSA" would degrade when we delete the other two components in turn.
Compared with the full model, the above experimental results primarily prove the superiorities of the designed components. Apparently, all the components can complement and reinforce each other, which further verifies the combined effects of the above semantic segmentation.

\subsection{Visualizations of Decoder Stages}
Compared with vanilla U-Net, the proposed TransAttUnet benefits greatly from the long-range feature dependencies and the global contextual information.
To further verify their abilities of semantic discrimination, we visualize feature maps of each decoder stage for both U-Net and TransAttUnet, as illustrated in Fig.~\ref{fig:9}.

Based on the comparative results, we can make the following observations:
1) It can be seen that the encoders of U-Net fail to make full use of the contextual information. With the guidance of the multi-level non-local attention mechanisms that effectively capture contexts, the obtained encoder features can provide more global semantic information to our decoders at lower stages, which can generate the most discriminative features.
2) Meanwhile, our TransAttUnet can make full use of the multi-scale contextual information to generate accurate predictions at later stages. As shown with the zoom-in patches of the deepest stage, the segmentation results of TransAttUnet are more detailed and reliable with a clearly discernible boundary.
Therefore, we can conclude that the semantic information learned by our method is more effective to improve the performance of medical image segmentation.

\section{Conclusion and Future Work}
\label{sec:5}
In this paper, we propose a novel Transformer based attention-guided U-Net called TransAttUnet, which concurrently incorporates multi-level guided attention and multi-scale skip connections into U-Net for improving the segmentation quality of biomedical images.
Specifically, the multi-level guided attention block is able to make full use of global contextual information by concurrently exploring long-range interactions and global spatial relationships between encoder semantic features.
Meanwhile, the multi-scale skip connection scheme can flexibly aggregate contextual feature maps from decoders of varying semantic scales to generate the discriminative feature representations.
Compared with previous advanced works, the proposed TransAttUnet benefits greatly from long-range feature dependencies and the multiscale contextual information,  which ensures the feature representations with semantic consistency. In this way, we can effectively mitigate the intrinsic limitations that occur in traditional U-shape architecture.
Extensive experimental results on different benchmark datasets demonstrate that the proposed TransAttUnet can achieve consistent performance improvements by integrating the above novelties.

Despite the fact that our study has some vital contributions, there are still several limitations. Admittedly, the proposed TransAttUnet heavily relies on the global self-attention mechanism and thus suffers large memory footprint and computation cost. Moreover,  the potential of Transformer for medical image segmentation remains incomplete and underutilized in the study of our work,  especially in the face of various biomedical images. Therefore, further improvements in these aspects will be investigated in our future work.



%


\ifCLASSOPTIONcaptionsoff
  \newpage
\fi



%
%
%
\bibliographystyle{IEEEtran} 
\bibliography{Benz_References}      

\end{document}